\documentclass[useAMS,usenatbib,usegraphicx]{mn2e}

\title[Bipolar jets growth and decline in Hen~3-1341]{Bipolar 
       jets growth and decline in Hen~3-1341: a direct link to 
       fast wind and outburst evolution\thanks{Based in part on observations 
       secured at the Observatoire de Haute-Provence within the OPTICON 2004 access program}}
\author[U. Munari, A. Siviero, A. Henden]{U. Munari$^{1}$\thanks{E-mail:
munari@pd.astro.it (UM); siviero@pd.astro.it (AS); aah@nofs.navy.mil (AH)}, 
A. Siviero$^{1}$\footnotemark[2] and A. Henden$^{2}$\footnotemark[2]\\
$^{1}$INAF-Osservatorio Astronomico di Padova, Sede di Asiago,
I-36012 Asiago (VI), Italy\\
$^{2}$Univ. Space Research Assoc./U. S. Naval Observatory,
P. O. Box 1149, Flagstaff AZ 86002-1149, USA}
\begin{document}

\date{Accepted ---. Received ---; in original form ---}

\pagerange{\pageref{firstpage}--\pageref{lastpage}} \pubyear{2005}

\maketitle

\label{firstpage}

\begin{abstract}
The appearance and disappearance of collimated bipolar jets in the symbiotic 
star Hen~3-1341 is reported and investigated. From modeling of the emission  
line spectrum it turns out that the accreting white dwarf in quiescence has  
$T_{\rm WD}$$\sim$1.2\,10$^5$~K and $R_{\rm WD}$$\sim$~0.14~R$_\odot$, for a 
luminosity of 3.8\,10$^3$~L$_\odot$, and it is stably burning hydrogen on the
surface at a rate of $\dot M_{\rm H}$$\sim$5\,10$^{-8}$\,M$_\odot$yr$^{-1}$,
feeding ionizing photons to a radiation bounded circumstellar nebula
extending for $\sim$17~AU. The WD underwent a multi-maxima outburst lasting from
1998 to 2004 during which its H-burning envelope reacted to a probable    
small increase in the mass accretion by expanding and cooling to $T_{\rm  
eff}$$\sim$1\,10$^4$~K and R$\sim$~20~R$_\odot$, mimicking an A-type giant
that radiated a total of $\sim$6\,10$^{44}$~erg, at an average rate of
$\sim$1\,10$^3$~L$_\odot$. Bipolar jets developed at the time of outburst maximum and
their strength declined in parallel with the demise of the fast wind from
the inflated WD, finally disappearing when the wind stopped halfway to
quiescence, marking a 1:1 correspondence between jets presence and feeding
action of the fast wind. The total mass in the jets was 
$M_{\rm jet}$$\sim$2.5\,10$^{-7}$~M$_\odot$ for a kinetic energy of 
$E^{kin}_{jet}$$\sim$1.7\,10$^{42}$($\sin i$)$^{-1}$~erg, corresponding to 
$\sim$0.3($\sin i$)$^{-1}$\% of the energy radiated during the whole outburst.
\end{abstract}


\section{Introduction}

\citet{b32}, hereafter TMM, discovered spectroscopically in Hen 3-1341 (=
V2523 Oph) one of the finest examples of highly collimated bipolar jets seen
in a symbiotic binary. The jets, with a projected velocity of
$|$$\Delta$RV$_\odot$$|$$\sim$820~km~sec$^{-1}$, were observed when the
system was in outburst at $V\sim$10.5 mag. No follow-up study monitored the
jet or outburst evolution.

Some symbiotic binaries are already known to present or have presented jets,
in the optical and/or radio: R~Aqr \citep{b05,b08}, CH~Cyg \citep{b28,b27},
MWC~560 \citep{b30,b22,b21}, RS~Oph \citep{b29}, Hen~2-104 \citep{b06} and
Hen~3-1341 (TMM). Other possible examples of jets or collimated mass
outflows from symbiotic binaries and related systems are StH$\alpha$~190
\citep{b20}, Z~And \citep{b04}, AG~Dra \citep{b15}, V1329~Cyg and HD~149427
\citep{b03}.

To correlate in symbiotic stars the appearance of jets with the system
properties is not an easy task. In fact, the ionized circumstellar gas is so
bright in the optical and ultraviolet as to prevent direct observation of
the central engine, which is believed to be an accreting white dwarf that in
many cases experiences quasi-stable surface H-burning \citep{b16,b26}. The
standard scenario for the production of jets (e.g. \citealt{b14}) involves
an accretion disk which is threaded by a vertical magnetic field and an
energy/wind source associated with the central accreting object.  A
widespread presence of accretion disks and magnetic fields in symbiotic
binaries is still matter of debate. Convincing detection of a magnetic field
in a symbiotic binary has been so far obtained only for Z~And \citep{b23},
in the form of a persistent $\leq$0.005~mag amplitude oscillation at 28~min
period interpreted as the spin period of a magnetic WD. The detection was
made possible by the outburst state of Z~And at the time of the
observations, which largely increased the optical brightness of the WD and
reduced that of the circumstellar nebular material by reducing the feeding
of ionizing photons. Direct evidences for accretion disks are missing, given
their low luminosity compared with the glare of the circumstellar nebular
material and the brightness of the central star (especially if burning
hydrogen at the surface). Indirect suggestions for the presence of disks are
generally based on the interpretation of the flickering so far detected in
$\sim$20\% of surveyed symbiotic stars (e.g. CH~Cyg: \citealt{b25}; T~CrB:
\citealt{b33}; MWC~560: \citealt{b31,b07}; RS~Oph and Mira A+B:
\citealt{b24}).

In this paper we present direct evidences that, at least in the symbiotic
binary Hen~3-1341, the appearance of bipolar jets is limited to the early
and brightest outburst phases, and that the jets are fed by the wind from
the outbursting component. When the wind quenches during the decline from
outburst maximum, the jets also vanish, in a 1:1 correspondence.

\begin{figure}
  \includegraphics[width=84mm]{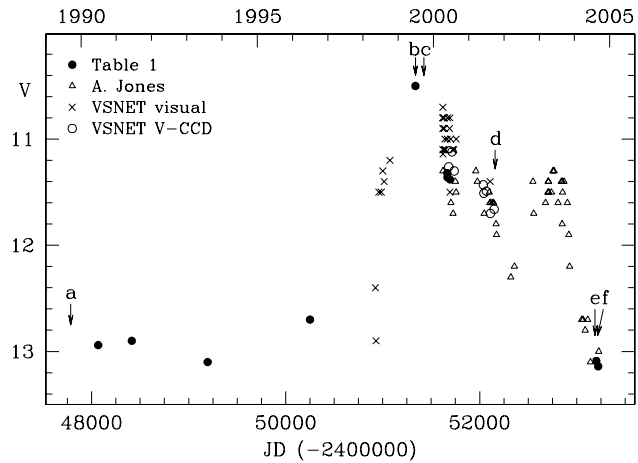} 
   \caption{The $V$-band light-curve of Hen~3-1341 over the last 15~yr
    showing the 1998-2004 outburst. The arrows point to the epochs of the
    spectra in Figure~3.}
  \label{lightcurve}
 \end{figure}

\section{Conditions in quiescence and outburst}

When TMM discovered the jets in the summer of 1999 they found Hen~3-1341 at
$V$=10.5, much brighter than on the Palomar charts ($B$=14.0), and concluded
the system was undergoing an outburst. Not much else is known about the
photometric and spectroscopic history of Hen~3-1341 or the reality and
characteristics of the outburst itself. A proper investigation is mandatory
for the obvious implications on the nature and evolution of the jets.

With the aim of recovering the photometric history of Hen~3-1341, we have ($a$)
obtained {\em UBV(RI)$_{\rm C}$JHK} photometry of Hen~3-1341 with the USNO
1.0m and 1.5m telescopes, ($b$) measured the brightness on photographic
plates found in the Asiago 67/92 Schmidt archive and on plates of the DSS-1,
DSS-2, GSC and Vehrenberg's photographic surveys, and ($c$) derived {\em
UBV(RI)$_{\rm C}$} magnitudes from published optical spectra calibrated into
absolute fluxes. These data are given in Table~1 together with values from
the 2MASS and DENIS survey, and {\em UBV(RI)$_{\rm C}$JHK} from the
symbiotic star photometric catalogs of \citet{b01} and \citet{b17}.
They are plotted in Figure~1, together with amateur estimates kindly
provided by Albert Jones (New Zealand) and VSNET organization, to document
the photometric evolution over the last 15~yr.

   \begin{figure*}
   \includegraphics[height=166mm,angle=270]{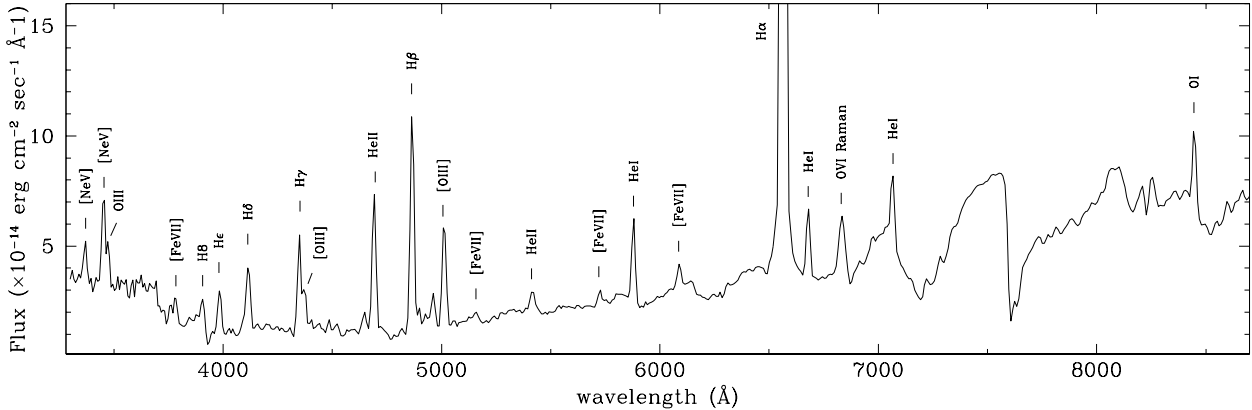}
   \caption{Low resolution spectrum of Hen~3-1341 in quiescence 
   on July 17, 2004}
   \label{spectrum}
   \end{figure*}
   \begin{figure*}
   \includegraphics[height=170mm,angle=270]{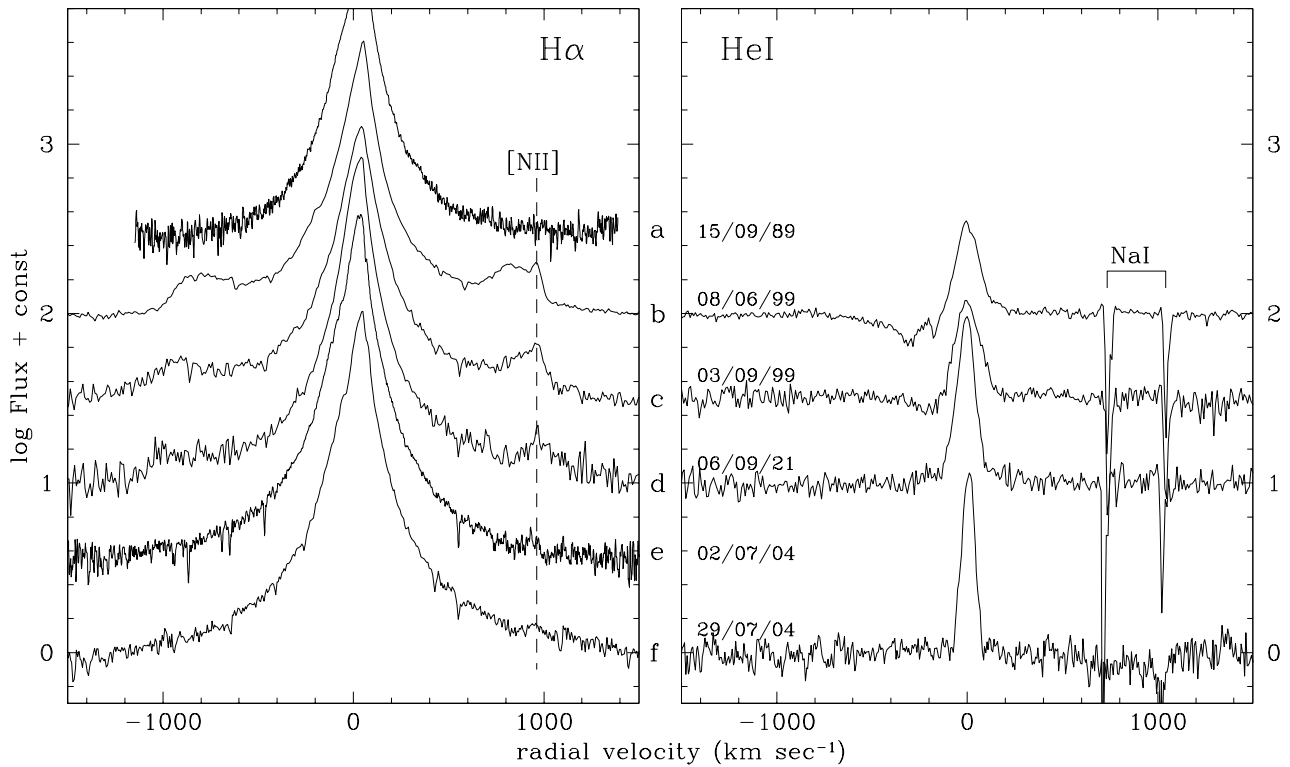}
   \caption{Evolution of H$\alpha$ and HeI~5876~\AA\ profiles between 1989
   and 2004. The profiles are plotted on a logarithmic flux scale to
   emphasize visibility of the jets in H$\alpha$ and the wind absorption
   signatures in HeI. $a$: from \citet{b34}; $b$,$c$,$d$,$f$:
   Echelle spectrograph at Asiago 1.82m telescope; $e$: ELODIE spectrograph
   at OHP 1.93m telescope.}
   \label{Halpha}
   \end{figure*}

Figure~1 reveals that Hen~3-1341 has undergone a large outburst lasting from
1998 to 2004, characterized by a rapid rise to maximum and a complex pattern
of minima and maxima during the decline phase. The outburst reached maximum
$V$ brightness right at the time TMM discovered the jets and found the high
ionization emission lines to have vanished and lower energy ones somewhat
weakened while an A-type continuum dominated the spectrum, like in a typical
symbiotic star outburst (cf. \citealt{b48}). The end of the outburst and return to quiescence by
summer 2004 is confirmed by the spectrum in Figure~2, that we obtained on
July 17, 2004 with the B\&C+CCD spectrograph of the 1.22m telescope operated
in Asiago by the Depart. of Astronomy of the Univ. of Padova. The spectrum
is characterized by strong emission lines of [FeVII], [NeV], HeII and OVI
(Raman scattered), identical to the 1993 pre-outburst quiescent spectrum
by \citet{b19}.

The emission line spectrum of Hen~3-1341 in quiescence both in the optical
(from Figure~2) and in the ultraviolet (the IUE spectrum of \citet{b10}) has
been modeled with the CLOUDY software package (see http://www.nublado.org/)
to get round estimates of basic physical parameters, assuming a simple
spherical symmetric gas distribution originating from the
$\rho$(r) $\propto$ r$^{-2}$ wind of the cool giant ionized by the radiation
from the WD companion. We found
the nebular material to be radiation bounded and to extend for $\sim$17~AU, for 
a total mass of $\sim$7\,10$^{-6}$~M$_\odot$ contained within.
The gas electronic temperature and density are $T_{\rm e}$$\sim$1\,10$^5$~K
and $N_{\rm e}$$\sim$2\,10$^8$~cm$^{-3}$ close to the ionizing source,
decreasing to $T_{\rm e}$$\sim$4\,10$^3$~K and $N_{\rm
e}$$\sim$6\,10$^5$~cm$^{-3}$ at the ionization boundary. The modeling
supports solar metallicity and chemical partition, although with Ne and N
apparently enhanced by $\sim$10$\times$ (which is what is expected when
ejecta or wind from the outbursting WD are mixed with the circumstellar material
originating from the M giant, cf. \citealt{b50},\citealt{b35}).
The central star has $T_{\rm WD}$$\sim$1.2\,10$^5$~K and $R_{\rm
WD}$$\sim$~0.14~R$_\odot$, for a luminosity of 3.8\,10$^3$~L$_\odot$, which
are consistent with those previously found by \citet{b10}. Such a
quiescent luminosity cannot be sustained by pure accretion since that would
require an implausibly large accretion rate of 1\,10$^{-6}$~M$_\odot$~yr$^{-1}$ on a
0.5~M$_\odot$ WD.

The quiescent luminosity is instead well explained by stable H-burning
conditions of accreted matter on the surface of the white dwarf. Because the
stably burning WD in Hen~3-1341 does not expand to large dimensions during
quiescence, the accretion rate must match the burning rate, or $\dot M_{\rm
H}$$\sim$5\,10$^{-8}$ ~M$_\odot$~yr$^{-1}$ which is in
line with the expected accretion rate for symbiotic stars (cf. \citealt{b49}).
From \citet{b12} numerical simulations, $T_{\rm
WD}$$\sim$1\,10$^5$~K and $R_{\rm WD}$$\sim$~0.1~R$_\odot$ correspond to a
stable H-burning 0.4~M$_\odot$ white dwarf. From Table~1 it is evident that
the white dwarf in Hen~3-1341 has been in that state since at least the time
of the first Palomar survey, 50 yr ago. This is in line with the theoretical
expectations for a 0.4~M$_\odot$ H-burning white dwarf, that predicts an
``on" period of $\sim$5000~yr, or ten times shorter if the mass of the 
envelope is reduced by a wind \citep{b13}.

\begin{table*}
 \centering
 \begin{minipage}{115mm}
   \caption{The table collects existing and new multi-band information on
   the photometric history of Hen~3-1341. Sources: $a$ = DSS-1, DSS-2, GSC,
   Vehrenberg's photographic surveys remeasured by us, $b$ = Asiago 67/92 cm
   Schmidt telescope archival plates, $c$ = \citet{b01}, $d$ = \citet{b16},
   $e$ = derived from covolution with band transmission
   profiles of the fluxed spectra in \citet{b10}, $f$ =
   similarly for \citet{b19} spectra, $g$ = 2MASS, $h$ = \citet{b32},
   $i$ = photometry with the USNO 1m telescope, $l$ = DENIS
   survey.}
 \begin{tabular}{rlllllcccc}
 \hline
 &&\\ 
 &
 \multicolumn{1}{c}{$U$}&
 \multicolumn{1}{c}{$B$}&
 \multicolumn{1}{c}{$V$}&
 \multicolumn{1}{c}{$R_{\rm C}$}&
 \multicolumn{1}{c}{$I_{\rm C}$}&
 \multicolumn{1}{c}{$K$}&
 \multicolumn{1}{c}{$J$$-$$H$}&
 \multicolumn{1}{c}{$H$$-$$K$}&
 \multicolumn{1}{c}{ref}\\ 
 &&\\ 
 26-04-1954 &        &      &       & 10.8 &       &      &      &      & $a$ \\ 
 01-07-1954 &        & 14.0 &       &      &       &      &      &      & $a$ \\ 
 14-07-1969 &        & 13.3 &       &      &       &      &      &      & $b$ \\ 
 16-07-1969 &        & 13.2 &       &      &       &      &      &      & $b$ \\ 
 06-04-1970 &        & 14.0 &       &      &       &      &      &      & $a$ \\ 
 02-06-1970 &        & 14.4 &       &      & 11.0  &      &      &      & $b$ \\ 
 01-07-1970 &        & 13.8 &       &      & 10.9  &      &      &      & $b$ \\ 
 $\sim$1981 &        &      &       &      &       & 7.58 & 0.98 & 0.34 & $c$ \\ 
 22-03-1982 &        & 13.0 &       &      &       &      &      &      & $a$ \\ 
 29-06-1987 &        &      & 12.5  &      &       &      &      &      & $a$ \\ 
 24-06-1990 & 12.72  & 13.73& 12.94 &11.82 & 10.82 &      &      &      & $d$ \\ 
 16-03-1990 &        &      &       &      &       & 7.66 & 1.06 & 0.31 & $d$ \\ 
 01-07-1991 & 12.6   & 13.7 & 12.9  &11.6  & 10.8  &      &      &      & $e$ \\ 
 09-03-1992 &        &      &       & 11.7 &       &      &      &      & $a$ \\ 
 27-07-1993 &        & 14.1 & 13.1  &12.0  &       &      &      &      & $f$ \\ 
 16-06-1996 &        &      & 12.7  &      &       &      &      &      & $a$ \\ 
 23-04-1998 &        &      &       &      &       & 7.48 & 0.87 & 0.41 & $g$ \\ 
 08-06-1999 &        &      & 10.5  &      &       &      &      &      & $h$ \\ 
 30-04-2000 & 10.97  & 11.82& 11.32 &10.65 & 10.03 &      &      &      & $i$ \\ 
 02-05-2000 & 11.00  & 11.91& 11.36 &10.74 & 10.07 &      &      &      & $i$ \\ 
 28-05-2000 & 11.08  & 11.90& 11.38 &10.70 & 10.09 &      &      &      & $i$ \\ 
 10-10-2000 &        &      &       &      & 10.11 & 7.50 &      &      & $l$ \\ 
 16-07-2004 & 13.15  & 13.84& 13.09 &11.72 & 10.72 &      &      &      & $i$ \\ 
 04-08-2004 & 13.22  & 13.91& 13.14 &11.74 & 10.74 & 7.56 & 0.94 & 0.36 & $i$ \\ 
 &&\\
 \hline
 \end{tabular}
 \end{minipage}
\end{table*}

To gain physical insight on the optical 1998-2004 outburst we have
integrated the energy between the outburst light-curve in Figure~1 and the
$V$$\sim$13.1 reference value for quiescence, to derive the total radiated
energy.  The color evolution of the outbursting component is missing for
lack of multi-band simultaneous observations. Color information is available
(cf. Table~1) only for the three observations in 2000, when the outbursting
component was roughly midway between outburst maximum and quiescence. For
sake of discussion, the corresponding $B$$-$$V$=0.52 is taken to represent
the average color of the outbursting component. To estimate the reddening we
have measured the equivalent width of interstellar NaI and KI absorption
lines in high resolution spectra of Hen~3-1341 that we have secured with
ELODIE spectrograph at the Observatoire de Haute-Provence (OHP) and the
Echelle spectrograph at the 1.82m telescope operated in Asiago by the
Italian National Institute of Astrophysics (INAF).  Following the
calibration of \citet{b18}, the reddening is found to be $E_{B-V}$=0.47 and
therefore the intrinsic average color for the outbursting component is
($B$$-$$V$)$_\circ$=0.05. The corresponding bolometric correction is $-$0.10
according to \citet{b02} and the spectral type is A2 following \citet{b09}.
Adopting the distance of 3.1~kpc derived by \citet{b10}, the total energy
radiated by the outburst turns out to be $\sim$6\,10$^{44}$~erg. It would
correspond to the potential energy released by the accretion onto a
0.5~M$_\odot$ white dwarf of $\sim$4\,10$^{-6}$~M$_\odot$ (at a mean rate of
$\sim$1\,10$^{-6}$~M$_\odot$yr$^{-1}$) which is implausibly large. In fact,
it is orders of magnitude larger than the typical mass dumped through an
accretion disk during outbursts in cataclysmic variables \citep{b35}, and
the required mass loss from the donor star would be orders of magnitude
larger than expected for the M2\,III non-variable companion, which did not
experience any major instability at the time of outburst onset (cf.
stable infrared magnitudes in Table~1).

The 6\,10$^{44}$~erg radiated during the outburst instead correspond to the
nuclear burning of $\sim$5\,10$^{-8}$~M$_\odot$ of Hydrogen, at an average
rate of $\sim$1\,10$^{-8}$~M$_\odot$yr$^{-1}$ or $\sim$1\,10$^3$~L$_\odot$.
This corresponds to about $\sim$1/4 of the luminosity in quiescence. This
suggests that the 1998-2004 event, that we previously called an ``outburst",
was actually a partial reprocessing into the optical of the radiation
flowing into the far UV during ``quiescence". In fact strong hydrogen and HeI emission lines 
remained always visible, in spite of a recombination time of the order of
$t_{\rm rec}$$\sim$1/($n_{\rm s}$$\alpha_{\rm B}$)$\sim$2~months even at the
outer boundary of the ionized gas region, indicating that not all UV radiation was
suppressed. The reprocessing
took place in the envelope of the WD that expanded and cooled to $T_{\rm
eff}$$\sim$1\,10$^4$~K and R$\sim$~20~R$_\odot$. The width of the stable
H-burning strip in the M$_{\rm WD}$, $\dot M_{\rm acc,WD}$ plane is a narrow
one, and minimal increases in $\dot M_{\rm acc,WD}$ triggers an
expansion and consequent cooling of the WD envelope (cf. \citealt{b13}). We are
therefore led to interpret the 1998-2004 ``outburst" as having been caused
by a temporary (and marginal) increase in the mass loss from the M2\,III
that gave no appreciable signal in the IR photometry of Table~1 but caused
$\dot M_{\rm acc,WD}$ to rise above the narrow equilibrium strip. The
surplus accreted material was partially burned and partially carried away by
the fast wind traced by the P-Cyg profiles of HeI lines in Figure~3. When
the WD envelope mass returned to equilibrium value, the envelope retraced to
small radius and high temperature and Hen~3-1341 resumed the photometric and
spectroscopic ``quiescent" appearance.

\section{Evolution of the jets and their feeding mechanism}

Following the discovery of bipolar jets by TMM in June 1999, we re-observed
Hen~3-1341 in high resolution at later dates. The results for H$\alpha$ and
HeI~5876~\AA\ profiles are shown in Figure~3 together with the original TMM
discovery spectrum and a pre-outburst spectrum obtained at ESO by
\citet{b34}. From Figure~3 it is evident how the bipolar jets were absent in
the quiescence {\em before} as well as {\em after} the outburst, and were
strongest at peak outburst optical brightness, declining in strength with
the outburst retracing from maximum. It is worth noticing the similarity
between the terminal wind velocity ($\sim$770~km~sec$^{-1}$) and jet
projected velocity ($|$$\Delta$RV$_\odot$$|$$\sim$820~km~sec$^{-1}$) at the
time of onset of the jets (June 1999), which is best appreciated in the left
insert of the bottom panel of Figure~1 of TMM. A most interesting comparison
is between jet strength and amount of mass loss by wind from the WD as
traced by the P-Cyg profiles of HeI lines in Figure~3. The correlation is a
very tight one. The jets were most prominent when the wind was strongest,
and declined in parallel with the decrease of wind intensity. To our
knowledge, these evidences are among the most direct and convincing proofs
of the `{\em energy/wind source associated with the central accreting
object}' postulated by \citet{b14} mechanism of jet production. We suggest
that the search for jets in symbiotic stars should be focused on the
outburst phases characterized by maximum wind intensity.

From the integrated flux of the jet features in the H$\alpha$ spectrum at
outburst maximum in Figure~3 (Jun 8, 1999), and assuming that all the
hydrogen within the jets is ionized and that the jets are optical thin in
H$\alpha$, a total mass in the jets of $M_{\rm
jet}$$\sim$2.5\,10$^{-7}$~M$_\odot$ is derived. The kinetic energy deposited in the
jets corresponding to the observed 820~km~sec$^{-1}$ projected velocity is
therefore $E^{kin}_{jet}$$\sim$1.7\,10$^{42}$($\sin i$)$^{-1}$~erg, where
$i$ is the unknown orbital inclination. This corresponds to $\sim$0.3($\sin
i$)$^{-1}$\% of the total energy radiated during the outburst.

Figure~3 suggests a small increase with time in the observed velocity
separation of the jets, amounting to
$|$$\Delta$RV$_\odot$$|$$\sim$820~km~sec$^{-1}$ on June 1999,
910~km~sec$^{-1}$ on Sept. 1999, and 1000~km~sec$^{-1}$ on Sept. 2001. 
Different explanations could be invoked - including a variable projection
angle of the jet axis on the line of sight as caused by a precession motion
- but there are not enough data to decide in favor of any of them.

Finally, we have also searched for a signature of rotation of a magnetic WD in
Hen~3-1341 by looking for coherent modulation in time resolved $B$-band
photometry that we secured in 2004 with the USNO 1.0m telescope (see
Figure~4). None was found, but this was an expected result, because with the
system in quiescence the circumstellar nebular emission is enormously
brighter that the WD in $B$ band. A more profitable search will have to wait
for the next outburst state of Hen~3-1341.

   \begin{figure}
   \includegraphics[width=84mm]{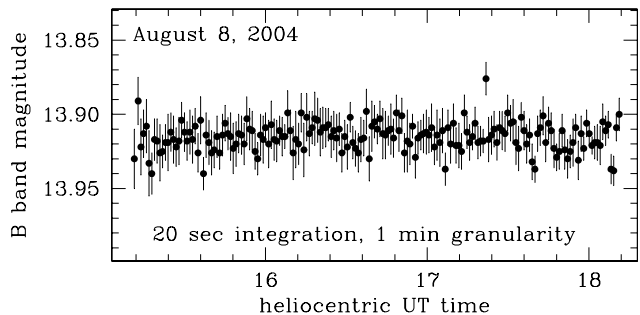}
   \caption{Time-resolved monitoring of Hen 3-1341 in $B$ band on August 8,
   2004 with the USNO 1m telescope, looking for flickering and signature of
   rotation of a magnetic WD.}
   \label{flickering}
   \end{figure}

%\section*{Acknowledgments}
%
%This work has been supported in part by Italian COFIN-2002
%and OPTICON 2004 grants.

\bsp

\label{lastpage}

\end{document}